\documentclass[sigconf,final]{acmart}

\usepackage{amsmath, amsthm}

\usepackage{pgfplots}
\usepackage{graphicx}
\usepackage{tikz}
  
\usepackage{lipsum}

\usepackage{xcolor}

\acmYear{2021}\copyrightyear{2021}
\setcopyright{rightsretained}
\acmConference[FAccT '21]{ACM Conference on Fairness, Accountability, and Transparency}{March 3--10, 2021}{Virtual Event, Canada}
\acmBooktitle{ACM Conference on Fairness, Accountability, and Transparency (FAccT '21), March 3--10, 2021, Virtual Event, Canada}
\acmPrice{15.00}
\acmDOI{10.1145/3442188.3445908}
\acmISBN{978-1-4503-8309-7/21/03}	


\newcommand{\pr}[1]{\mathtt{Pr}[#1]}

\title{A Bayesian Model of Cash Bail Decisions}
\author[1]{Joshua Williams}
\affiliation{Computer Science Department, \and Carnegie Mellon University \and \textbf{jnwillia@cs.cmu.edu}}

\author[1]{J. Zico Kolter}
\affiliation{Computer Science Department, \and Carnegie Mellon University \and Bosch Center for AI \and \textbf{zkolter@cs.cmu.edu}}

\begin{document}


\begin{abstract}
The use of cash bail as a mechanism for detaining defendants pre-trial is an often-criticized system that many have argued violates the presumption of ``innocent until proven guilty.'' Many studies have sought to understand both the long-term effects of cash bail's use and the disparate rate of cash bail assignments along demographic lines (race, gender, etc). However, such work is often susceptible to problems of infra-marginality -- that the data we observe can only describe average outcomes, and not the outcomes associated with the marginal decision.
In this work, we address this problem by creating a hierarchical Bayesian model of cash bail assignments.  Specifically, our approach models cash bail decisions as a probabilistic process whereby judges balance the relative costs of assigning cash bail with the cost of defendants potentially skipping court dates, and where these skip probabilities are estimated based upon features of the individual case. We then use Monte Carlo inference to sample the distribution over these costs for different magistrates and across different races.  We fit this model to a data set we have collected of over 50,000 court cases in the Allegheny and Philadelphia counties in Pennsylvania.  Our analysis of 50 separate judges shows that they are \emph{uniformly} more likely to assign cash bail to black defendants than to white defendants, even given identical likelihood of skipping a court appearance.  This analysis raises further questions about the equity of the practice of cash bail, irrespective of its underlying legal justification.  
\end{abstract}

\maketitle


\section{Introduction}


In the criminal justice system, cash bail is the often-used practice of requiring that defendants pay a monetary fee in order to be released pretrial, which is returned upon making all necessary court appearances. The inability to pay the set bail amount results in defendants being detained in jail for no reason other than their inability to pay an often arbitrarily large fee. The practice has come under a great deal of criticism, with many experts arguing that it violates the underlying notion of ``innocent until proven guilty.''  Indeed, such pretrial detention, for those unable to afford bail, or for those who need to seek commercial bail bonds in order to secure their bail, is one of the main drivers of the United States' growing incarceration rate~\cite{keller2017money}. 

The effects of pretrial detention are larger than just ensuring a defendant appears at their trial; there are lasting, devastating consequences on individual, family, and job security~\cite{heaton2017downstream}. Pretrial detention has been found to coerce defendants to plead guilty, regardless of reality, for the opportunity to return home, rebuild their lives and prepare for trial~\cite{sacks2012pretrial}.  Moreover, several other studies have found that pretrial detention increases the likelihood of conviction, increases the chance that a defendant will receive a mandatory minimum sentence if charged, and increases the defendant's ultimate sentence~\cite{stevenson2017pretrial, didwania2020immediate}. The use of cash bail alone has also been suggested to make defendants more likely to recidivate~\cite{gupta2016heavy}, i.e. it acts counter to the belief that our justice systems are rehabilitative. Many jurisdictions have taken steps to address the concerns of cash bail decisions, including  changes in the necessary requirements for a magistrate to issue cash bail, the use of pretrial risk assessments with the intent to curb human biases in the decision-making process, or altogether stopping cash bail practices~\cite{whatshappening2020}. 

In addition to the studied effects of cash bail, a further issue arises in considering the disparate rates of its use across groups (ie. race, gender, sexual identity, etc). Many prior studies describe this disparity by demonstrating the different rates of cash bail assignments with respect to race, gender, wealth, etc. However, as has been recently highlighted by the statistics community in many other settings, such benchmark tests are susceptible to the problem of infra-marginality. Due to differences in the unobserved marginal distributions, observing average outcomes does not yield enough information in order to determine whether a particular decision-making process is biased ~\cite{simoiu2017problem,ayres2002outcome, neil2019methodological}. The outcomes that we observe may or may not be a necessary result of the current reality. 

In this paper, we propose a stylized model of bail decisions, which directly estimates how cash bail assignments result from different (abstract) decision-making costs associated with defendants appearing in court -- i.e., we assume that cash bail is assigned if the ``expected cost'' of assigning cash bail is lower than the ``expected cost'' of the defendant potentially failing to appear for a court date (FTA). This captures a loose categorization of the National Association of pretrial Services Agencies (NAPSA) 2020 guiding principles~\cite{napsa2020guiding} for pretrial decision making in that: 1) the goal of bail setting is to find a bail amount and bail type that maximizes a defendant's likelihood of pretrial release, while ensuring a defendant's court appearance; 2) these bail decisions and amounts should not impose a disparate or discriminatory outcome based on race, gender, sexual identity and other legally protected attributes; 3) pretrial detention should be limited only to those defendants who pose an unmanageable risk to public safety. Under the proposed model, we infer a latent variable which captures the cost that each magistrate believes that society bears by a defendant being unable to pay the set cash bail versus the cost of this defendant failing to appear for a court date (itself based upon a linear model of probability of an FTA as a function of the covariates describing the particular case in question). Because we specifically allow these probabilities to \emph{depend} upon group characteristics such as the race of the defendant, the model separates out the relative decision-making costs of bail from the actual probabilities involved.  This enables the model to deal with the infra-marginality problem to some degree, and we estimate the mechanism behind a magistrate's cash bail decisions \emph{even accounting for potential differences in FTA rates amongst different groups.}

We estimate these costs by creating a dataset of $89,751$ court dockets from Philadelphia and Allegheny counties in Pennsylvania from the years 2016 to 2019. We then fit the model parameters for the 50 most active magistrates within our dataset. Here, we found that \emph{nearly every one of the 50 most active magistrates shows a greater willingness to assign cash bail to Black defendants than to White defendants, even given equal probabilities of skipping trial}. In other words, for two defendants with the same relative probability of skipping their court appearance, judges were uniformly more likely to assign cash bail to a Black defendant.  This is particularly striking given the fact that some of the simpler baseline analyses (such as simply comparing bail assignment rates for black and white defendants) show more similar treatment for each group.  From a more algorithmic perspective, we believe our methodology also presents a valuable contribution to work on addressing infra-marginality problems via Bayesian modeling.  While past work has addressed similar questions, to the best of our knowledge ours is the first to simultaneously estimate a full parametric predictive model of relevant outcomes that includes group identifying components, while simultaneously attempting to estimate the relative costs of decisions based upon these groups.  Thus, we believe that the current work presents both an important study from a societal perspective, and a methodological advance of interest to the broader algorithmic community.
	 

\section{Background and Related Work}

In this section, we introduce the pretrial process, as it relates to bail decision-making in Allegheny County and Philadelphia County, and address prior analysis for this aspect of the judicial process.  

\subsection{The Process of Cash Bail}

In Philadelphia County, after an arrest, typically a video conference is set up between a magistrate and a defendant, where the magistrate will assign a bail type and bail amount in a hearing that generally lasts one to three minutes~\cite{nworah2017cost,aclu2020challenge}. Each shift of bail hearings takes place for approximately four hours every day, with one of Philadelphia's six arraignment court magistrates presiding over each shift~\cite{philadelphia2018report}. Pretrial processes work much the same in Allegheny county, where one of 46 magisterial district judges will preside over a preliminary arraignment in which a defendant's bail is set. Bail need not be set only at the preliminary arraignment, it may also be set during a preliminary hearing or bail hearing. However, in most criminal cases, it is set at this preliminary arraignment. 
	
Of the bail types assigned, a magistrate has the option under the law~\cite{pacode524} to choose one of the following:
\begin{enumerate}
	\item Release on recognizance (ROR) -- a written agreement from a defendant to appear on their court date. This agreement may come with additional stipulations, but does not require any monetary commitment.
	\item Release on nonmonetary conditions -- a magistrate may release a defendant on nonmonetary bail if they believe that set conditions such as restricting travel or requiring the defendant to report in are sufficient for their court appearance.
	\item Release on nominal bail -- a magistrate may release a defendant on nominal bail by requiring a small amount of money (eg. one dollar) and another entity (a person or organization) to act as surety.
	\item Release on unsecured bail -- a magistrate may release a defendant on unsecured bail by requiring a written agreement that the defendant becomes liable for the set amount of money in the event of non-appearance. This does not require any money or other form of security in order to be released.
	\item Release on monetary bail -- a magistrate may release a defendant on cash bail if they believe that a defendant will be unlikely to comply with the conditions of release without an immediate guarantee. In Allegheny County, the bail authority may require that individual to pay no more than $10\%$ of the full bail amount in order to secure their release. If a defendant is unable to pay, they will be detained until trial~\cite{allecntycode528}. In Allegheny County, cash bail should not be set unless a magistrate investigates the defendant's financial background and finds that the set amount is reasonable~\cite{pacode528}.
\end{enumerate}
	
\subsection{Implications of Cash Bail Assignments}

	If a defendant is detained pretrial because the set bail was not reasonable then there can be serious consequences, both inside and outside of the case. In a qualitative assessment of 23 interviews among defendants who pled guilty, \citet{euvrard2017pre} found, among other effects, that pretrial detention acts as a coercive mechanism in which by pleading guilty, the defendant can avoid jail time, whereas by pursuing a trial, the defendant will have to wait in jail for an undetermined amount of time. Similarly, some interviewees acknowledged that the jail-time offered by accepting a plea deal may be equivalent to their time spent in jail by waiting for trial. Pleading guilty often leads to the minimum amount of time spent in jail, regardless of innocence or guilt. 
	
	Further quantitative studies have confirmed similar effects. In a study of New York's pretrial outcomes, by regressing conviction and pretrial detainment on defendants' relevant features, \citet{leslie2017unintended} found a positive correlation between pretrial detainment and conviction. In other words, the authors find that the more likely a judge is to detain an individual pretrial, the more likely that this individual pleads guilty or is convicted of a crime. In a similar analysis, \citet{dobbie2018effects} use a measure of judge leniency, based on the residuals of release decisions in a probit model, in order to estimate the influence of the assigned magistrate on pretrial release and its subsequent effects on future crime, court appearances, and employment. They find that while pretrial release increases the likelihood of failing to appear, release both decreases the likelihood of rearrest over the next two years and increases the likelihood of having any income over the next two years.
	
	In particular, such models generally follow the rationale that the impact that the presiding authority has on a case can be isolated and that we can estimate regression coefficients for outcomes in ways that glean new insights, as in the example below.  For example, \citet{gupta2016heavy} propose a regression model
\begin{align}
	&Z_{ictjo} = \frac{1}{n_{ctjo}-1} \sum_{k \neq i} \Big( \mathrm{Bail}_{kctjo} \Big) - \frac{1}{ n_{ct} - 1 } \sum_{k \neq i} \Big( \mathrm{Bail}_{kct} \Big) \\
	&\mathrm{Outcome}_{ictjo} = \alpha + \beta \mathrm{Bail}_{icto} + \delta X'_{icto} + \epsilon_{ictjo} \\
	&\mathrm{Bail}_{ictjo} = \alpha + \gamma Z_{ictjo} + X' \zeta + \nu_{ictjo}
		\label{leniencyreg}
	\end{align}
	where $\mathrm{Bail}_{ictjo}$ is an indicator for whether or not cash bail was set for defendant $i$ with offense $o$ in court $c$ for year $t$ by judge $j$ and $X'$ are defendant controls, such as age, race, gender, charges, priors, etc. $Z_{ictj}$ is a measure of judge leniency, calculated as the difference in leave-one-out means of a judge's rate of cash bail assignments compared to the rate of cash bail assignments for court $c$. A litany of prior literature~\cite{gupta2016heavy,arnold2018racial,heaton2017downstream,donnelly2018downstream,stevenson2018distortion} uses such methods based on probit/logistic regression models with or without residual measures of judge leniency/severity in order to model how our observed pretrial data influences defendant outcomes after their preliminary arraignments/bail hearings, with most work coming to similar conclusions.
	
\subsection{Disparate Rates of Cash Bail Assignments}

	Due to the harmful long-term impact of cash bail use, further work has sought to verify whether or not magistrates assign cash bail uniformly among demographic groups. For cases in which a judge assigned cash bail, prior work~\cite{demuth2003racial, sacks2015sentenced, vaughn2019new, aclu2019punishing, steffensmeier2001ethnicity} has reviewed pretrial decisions for instances of judge bias and found that across the state of Pennsylvania, the proportion of Black defendants who are assigned cash bail is significantly greater than the proportion of their non-Black counterparts who were assigned cash bail, with additional work~\cite{santhosh2020investigating} finding that even after controlling for the types of crimes, judges still assign cash bail to Black and White defendants differently for the same crime in $20\%$ of cases. 
	
  Furthermore, a 2019 investigation by the American Civil Liberties Union (ACLU)~\cite{aclu2019punishing} found that cash bail was assigned in as many as $25\%$ of cases, with some magistrates setting cash bail in up to $57\%$ of cases. In Philadelphia, a 2018 review by the ACLU found that upwards of $43\%$ of defendants were assigned cash bail~\cite{aclu2018request}. Other studies~\cite{ouss2020bail,santhosh2020investigating} have also found cash bail assignment rates ranging from $40\%$ to $70\%$ over all magistrates.
	
	Beyond biases along racial lines, a 2017 class action suit brought against Harris County, Texas~\cite{odonnellvharris} argued that the county had been using money bail on misdemeanor defendants who could not pay. The district court, who's ruling was later affirmed by the Fifth Circuit Court of Appeals, ``concluded [that] Hearing Officers were aware that, by imposing a secured bail on indigent arrestees, they were ensuring that those arrestees would remain detained.''~\cite{odonnellvharris}. In essence, from jurisdiction to jurisdiction, we see trends that show that cash bail has been used as a barrier to justice for individuals of a certain race or financial status.
	
\subsection{The Problem of Infra-Marginality in Cash Bail Assignments}
	
Among the studies on both the disparate rate of cash bail assignment and the downstream effects of cash bail, there has been a growing emphasis that the methods employed in these studies may not be able to show the entire story~\cite{neil2019methodological}. Specifically, such methods are susceptible to the problem of infra-marginality. As coined in~\cite{ayres2002outcome}, the infra-marginality problem notes that our observed data is only able to provide average outcomes and not the outcomes associated with the marginal decisions. This problem has the potential to undermine analyses that are based solely on the observed information, causing such work to be unable to provide a complete picture of the underlying reasons for observed disparities. Beyond the infra-marginality problem, studies of bail decisions are also susceptible to Simpson's paradox, where the marginal association between categorical variables changes after controlling for one or more other variables. 
	
\begin{figure}
		\includegraphics[scale=0.17]{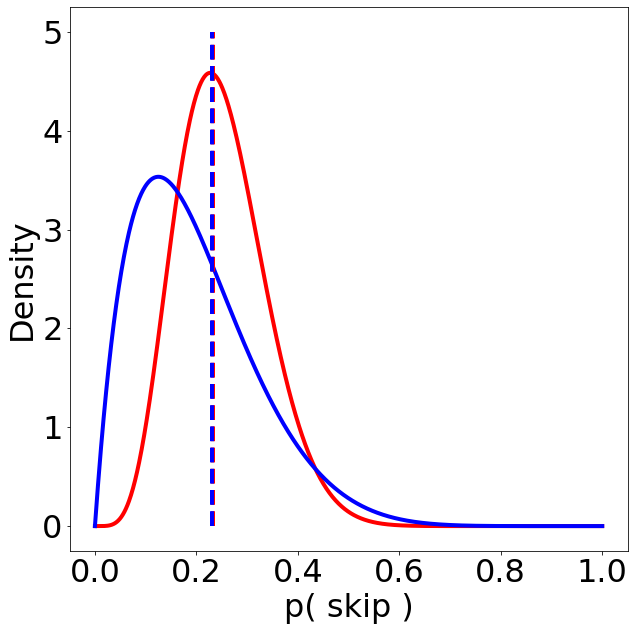}
		\includegraphics[scale=0.17]{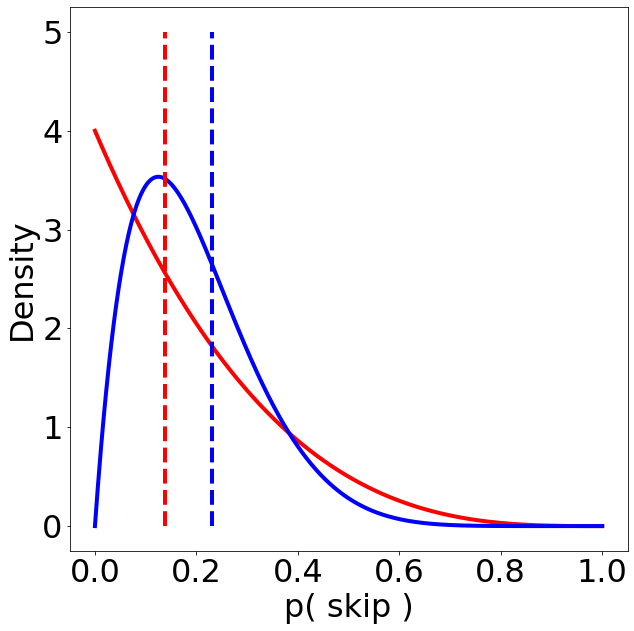}
		\caption{How the infra-marginality problem may manifest in bail analyses. (Left) Example Beta Distributions of FTA probabilities: red=Beta(6,18), blue=Beta(2,8). (Right) Example Beta Distributions of FTA probabilities: red=Beta(1,4), blue=Beta(2,8);  In both cases, thresholds are set so that the proportion above the threshold are $0.55$ and $0.35$ for red and blue groups respectively. Additionally thresholds are set so that expectation above the threshold are $0.311$ and $0.335$ respectively. Outcome statistics remain unchanged despite the threshold's reflecting different policies by a judge.}
		\label{inframargfig}
	\end{figure}

	 As an illustrative example, suppose that there exist two groups with different probabilities of missing their court dates, and that there exists some threshold where a judge assigns cash bail to all defendants whose court failure to appear (FTA) probability exceeds this threshold. We may observe that one group is assigned cash bail at a higher rate than a second and we may observe that under a judge's cash bail assignments, one group is less likely to miss a court appearance than another, we may reasonably conclude that there is bias in a judge's decisions. However,  without prior knowledge of the underlying distributions and the marginal decision threshold, there exist cases in which these seemingly biased outcomes occur without an bias on the part of the judge. For example, figure \ref{inframargfig} shows two hypothetical cases in which both the rate of cash bail assignment for the red and blue groups (0.55, 0.35) and the expected FTA rate for those who were assigned cash bail for each group (0.311, 0.335) does not change. 

	These observed statistics tell us that a judge is assigning cash bail to the red group at a higher rate, and that the expected FTA rate for defendants in the red group who are assigned cash bail is also lower than the expected FTA rate for those members in the blue group who receive cash bail. This may be indicative of some bias on the part of the judge. However, if we consider the judge's bail assignment thresholds, the first case shows that the judge is using a consistent decision threshold, irrespective of group membership. The disparate rates of cash bail assignment are not based on bias on the part of the judge, but a result of each group's marginal FTA rate. In the second case, the rate of cash bail assignments and expected FTA rates under cash bail assignments remain the same as in the first case, however, the decision threshold shows that the judge is assigning cash bail to members of the red group who have a lower probability of failing to appear than the blue group. Here, the differences in cash bail assignment rates and FTA rates are actually a result of bias on the part of the judge. The problem of infra-marginality lies in such cases, where, due to the fact that we cannot observe the marginal distribution of FTA probabilities, the observed data is unable to differentiate between the regime where a judge favors one group and one where judges are unbiased with respect to group membership. For a more detailed example, see~\cite{simoiu2017problem}.
	
	\citet{ayres2002outcome} explicitly argues that the bail-bond setting avoids the infra-marginality problem. The author argues that that the problem in cash bail assignment is non-dichotomous, as judges also assign a continuous variable, bail amount. By assigning bail amounts in a way that changes individual FTA probabilities, we effectively marginalize the distribution over all covariates, so that the observed outcomes are outcomes on the true marginal distribution. However, while the assigned bail amount is continuous, the decision to assign cash bail in the first place is dichotomous; judges choose to use cash bail in lieu of four other options. Before cash bail amounts are considered, we are already in the setting in which Ayres argues that the problem of infra-marginality undermines outcome tests.
	
	Even considering the setting where we condition on defendants who were assigned cash bail, the notion that bail amounts allow us to avoid this problem requires the assumption that a judge acts on full knowledge of a defendant's ability to pay, which is unlikely given the short (1-3 minute) nature of preliminary hearings. If a judge assigns cash bail at an amount that a defendant cannot pay, then we are still effectively marginalizing over two distinct subgroups, those defendants detained pretrial and those defendants released on cash bail. Such observations are then subject to Simpson's paradox, in which the trends change after again conditioning on those who are released pretrial on cash bail.
	
	
	Moreover, such models as equation \eqref{leniencyreg}, that emphasize a residual measure of judge leniency, are also susceptible to infra-marginality effects. For instance, the model above can be simplified to 
	\begin{small}
	\begin{align}
		\mathrm{Outcome}_{ictjo} &= \alpha + \beta \gamma Z_{ictjo} + ( \delta + \zeta ) X'_{icto} + \epsilon_{ictjo} + \nu_{ictjo}
		\label{simplereg}
	\end{align}
	\end{small}
	As judge leniency measures themselves are the difference of successive benchmark tests, emphasizing measures of judge leniency introduces a new component that suffers from the problem of infra-marginality. 
	While such analyses provide valuable insight, finding ways to address this shortcoming is valuable for a full understanding of such processes.  
	
	This issue of infra-marginality has been directly addressed in a different setting~\cite{simoiu2017problem,pierson2018fast} in an analysis of nationwide traffic stops, with a hierarchical Bayesian model that fits a threshold to the marginal probability distribution of contraband possession among groups. Officers are not expected to search a vehicle unless the likelihood of contraband possession is above this fit threshold. This setup avoids the issue of infra-marginality, by not directly basing indicators of bias on the observed outcomes, but on the latent variables that describe these outcomes.
	
	We follow a similar problem setup in this work, in which we posit a hierarchical Bayesian model of cash bail assignments based on the a group's FTA (failure to appear) rate and their pretrial release rate. The FTA and release probabilities are then used to fit a latent variable, based on a magistrate's perceived societal cost of members of a group failing to appear for their court dates or being unable to post their bail.  In our approach, the latent variables describing economic costs \emph{and} the parameters of the predictive model are simultaneously estimated via MCMC sampling.  Thus, the model presented here makes contributions on both the algorithmic and societal levels, in two ways: 1) providing a new example in which Bayesian modeling may be able to address problems of infra-marginality in complex decision-making systems 2) adding to existing evidence within the conversation regarding group-based biases within the United States judicial system.

\section{A Bayesian Hierarchical Bail Assignment Model}
	In order to mitigate the infra-marginality effect, we take a similar approach as \citet{simoiu2017problem} in creating a generative model that compares the probability distributions of different groups, and finds outcomes which are dependent on both the probability distributions and a latent variable that describes the underlying beliefs for each magistrate. Here, we assume a simple process bail assignment. We base this analysis on two major assumptions for the behavior of judges: 1) That cash bail should be a means of decreasing the likelihood that a defendant will fail to appear for a court appearance 2) Judges act rationally, using cash bail to minimize the balance in costs between assigning cash bail and defendants missing a court date. In other words, we disregard the possibility that cash bail is used in a punitive manner by any judge, and assuming that judges assign cash bail in order to minimize some (abstract) cost. While these assumptions may not always be satisfied in practice \cite{odonnellvharris}, we are effectively assuming the commonly understood purpose of bail, that bail is simply meant to ensure defendants will appear for trial and all pretrial hearings for which they must be present \cite{bar2019how} and that (as per law) judges apply the least restrictive conditions that will reasonable ensure a defendant's attendance at future court proceedings and enhance public safety \cite{house2017pretrial}.
	
		Our model then follows the rationale that whenever a case comes before the presiding authority, all available information, $x \in \mathbb{R}^{d}$, such as severity of the charged crime and number of prior crimes, dictates the probability of this defendant not appearing on their trial date. We model this probability of failing to appear as a logistic model with a Normal prior on the weights,
		\begin{equation}
		\theta_{\mathrm{fta}} \sim \mathcal{N}( 0, 2 I ).
		\end{equation}
		To model the effect of cash bail on FTA rates, we introduce an additional coefficient with prior $\theta_{b_1} \sim \mathcal{N}_{+}( \sigma_{b} )$, where $\mathcal{N}_{+}$ is the half-normal distribution (making the reasonable assumption that adding cash bail should not \emph{increase} the probability of failing to appear for a court date). Thus, the probability of missing a court date is given as,
		\begin{equation}
			\pr{ \mathrm{fta} | b } = \sigma( \theta_{\mathrm{fta}}^{T} x - \theta_{b_1} b + \theta_{0} ),
		\end{equation} 		
		where $\sigma$ is the sigmoid function, $b$ is the indicator for whether or not cash bail was assigned and $\theta_{0} \sim \mathcal{N}( 0, 2 )$ is a bias term, again with a Normal prior.
		
		Secondly, we consider the case where the presiding authority may be assigning the bail amount in excess of what the defendant is able to reasonably pay. We model the probability of a defendant's release for any type of bail by a second logistic model with a Normal prior
		\begin{equation}
		\theta_{\mathrm{release}} \sim \mathcal{N}( 0, 2 I )
		\end{equation}
		and the probability of being released pretrial given by
		\begin{align}
			\begin{split}
				\pr{ \mathrm{release} | b } &= \sigma( \theta_{\mathrm{release}}^{T} x - \theta_{b_2} b + \theta_{1} )
			\end{split},
		\end{align}
	where, $\sigma$ is the sigmoid, $\theta_{b_2} \sim \mathcal{N}_{+}( \sigma_{b} )$, and $\theta_{1} \sim \mathcal{N}( 0, 2 )$ is bias term.
	
	After estimating these probabilities, each magistrate determines whether to set cash bail based on a tradeoff between the probability of the defendant appearing if released pretrial with/without cash bail, and the probability of them being able to post their bail if cash bail is assigned. Each magistrate has a measure of societal harm induced by a defendant not coming to their trial. We represent this cost of not appearing, as the latent variable $\tau_{1} \sim \mathcal{N}_{+}( \sigma )$. 
	
	Should the presiding authority believe that cash bail is an appropriate assignment for a defendant, they legally must still ensure that the bail amount is reasonable. While reasonable may mean different things to different magistrates, we represent the cost of the defendant facing pretrial incarceration (i.e., having cash bail imposed and being unable to pay) via the latent variable $\tau_{2} \sim \mathcal{N}_{+}( \sigma )$.

\subsection{Cost Model}
	
	
	In the model here, these underlying beliefs, $\tau_{1}, \tau_{2}$, determine whether cash bail is an effective tool for bringing a specific defendant to their trial date. If a defendant is likely to not appear for a court appearance, regardless of the type of bail set, instead of setting cash bail, a magistrate will opt to deny bail to the defendant. However, if cash bail acts as a deterrent for a missed court appearance and the defendant can afford it, then the magistrate will instead choose to set cash bail. Alternatively, if the defendant cannot afford bail, and they are still unlikely to miss their court date under non-monetary release, the magistrate will choose not to use cash bail.  
	
	We represent this dilemma as choosing the option that minimizes the cost of setting cash bail,
	\begin{equation}
	\tau_{1} \pr{ \mathrm{fta} | \mathrm{cash\ bail} = 1} + \tau_{2} ( 1 - \pr{ \mathrm{post} | \mathrm{cash\ bail} = 1} )
	\end{equation} 
	or the cost of not using cash bail,
	\begin{equation}
	\tau_{1} \pr{ \mathrm{fta} | \mathrm{cash\ bail} = 0} + \tau_{2} ( 1 - \pr{ \mathrm{post} | \mathrm{cash\ bail} = 0 } )
	\end{equation}
    (Note if cash bail is not used, then the defendant posts bail with probability 1). For convenience in notation, let $p_{b}[ Y ] = \pr{ Y | \mathrm{cash\ bail} = b }$. 
	In this model cash bail is assigned if,
	\begin{align}
		\begin{split}
		\tau_{1} p_{1}[ \mathrm{fta} ] + \tau_{2} ( 1 - p_{1}[ \mathrm{release} ] ) \leq \tau_{1} p_{0}[ \mathrm{fta} ] 
		\end{split}
		\label{basemodel}
	\end{align}
	
	Rather than estimating both $\tau_{1}, \tau_{2}$, we simplify this equation to have a single latent variable, $\tau = \frac{\tau_{1}}{\tau_{2}}$, $\tau \sim C_{+}( 0, 1 )$, where $C_{+}$ is the Half-Cauchy distribution. This distribution is chosen by considering the case where $\tau_{1}, \tau_{2}$ are both Half-Normal random variables with the same variance. The ratio distribution, $\frac{\tau_{1}}{\tau_{2}}$ is itself a random Half-Cauchy variable with location $0$ and scale $1$.
	
	  Simplifying equation \eqref{basemodel}, cash bail is set if,
	\begin{align}
		\begin{split}
		\tau ( p_{0}[ \mathrm{fta} ] - p_{1}[ \mathrm{fta} ] ) - ( 1 - p_{1}[ \mathrm{release} ] ) \geq 0
		\label{taumodel}
		\end{split}.
	\end{align}
	We would like to enforce this constraint by modeling the observed cash bail as a Bernoulli random variable, with probability, 
	\begin{align}
		\begin{split}
		\hat{p}( \mathrm{cash\ bail} ) = \sigma\Big( \tau ( p_{0}[ \mathrm{fta} ] - p_{1}[ \mathrm{fta} ] ) - ( 1 - p_{1}[ \mathrm{release} ] ) \Big),
		\end{split}
	\end{align}
	 where $\sigma$ is the inverse logit function. In doing so, we capture the notion that as the cost of setting cash bail increases, the probability of setting cash bail falls. 
	
	\begin{figure}
		\centering
		\input{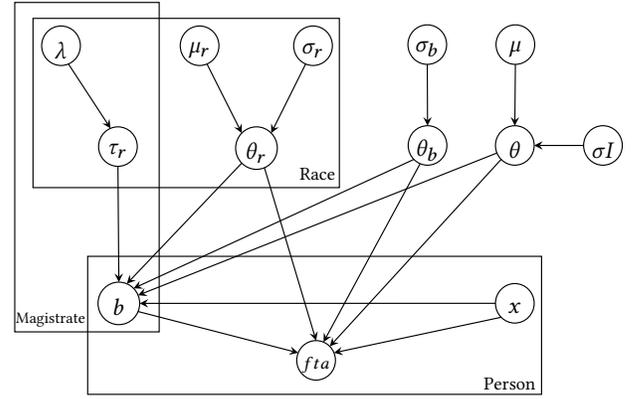}
		\caption{Generative model describing how magistrates set cash bail and how these decisions influence probabilities of failing to appear. Not pictured here is the analogous model for individual probabilities of pretrial release; both generative models use the same $\tau_{r}$ parameter from the magistrate plate. }
		\label{Diagram1}
	\end{figure}	
	
	While ideally it captures our constraint, in its current form, the inverse logit alone cannot model the probability of bail decisions due to the bias being governed by $( 1 - p_{1}[ \mathrm{cash\ bail} ]$. This bias is limited to the range $[0,1]$, which in turn would cause the output probabilities to range from $[ \frac{1}{1+e}, 1]$, instead of $[0,1]$. In order to address this, we rescale the sigmoid outputs in order to force the probability of assigning cash bail back into the $[0,1]$ range. The marginal probability of being assigned cash bail is just given by,
	\begin{equation}
		p( \mathrm{cash bail} ) = \frac{ \hat{p}( \mathrm{cash bail} ) - \frac{1}{1+e} }{ 1 - \frac{1}{1+e} }
		\label{sigmarescale}
	\end{equation}
	Our final hierarchical Bayesian model is shown in Figure \ref{Diagram1}.
	

	\subsection{Bail Amounts}
	Finally, it is worth noting that while the set cash bail amount is available in our data, we do not include it in this model. In order to estimate the cost of setting/not setting cash bail, we need counterfactual information on bail assignments. The probability of failing to appear for a court appearance is designed so that the counterfactual cash bail assignment is easily computable, however, due to the inherent stochasticity of the bail amounts that judges choose, we found that a counterfactual assignment of cash bail will not provide a reasonable estimate of what a defendant's bail amount would be had the judge set cash/non-monetary bail. 
	
	\begin{figure}
		\includegraphics[scale=0.25]{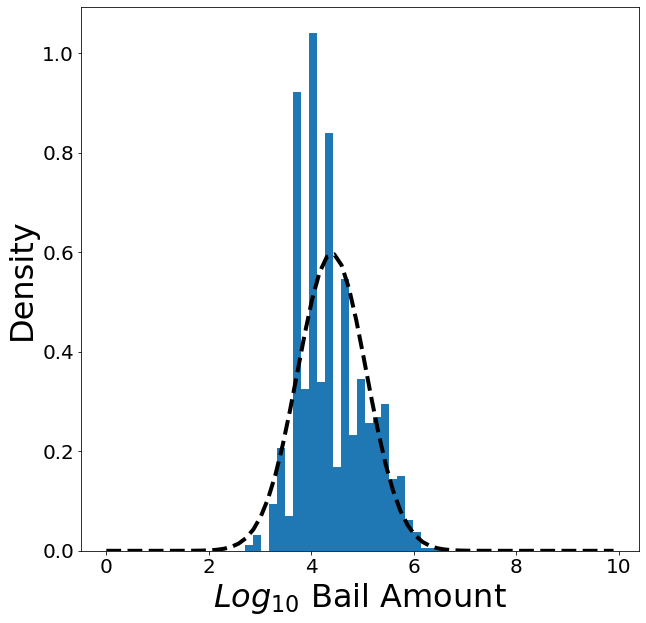}
		\caption{A histogram and the fitted normal distribution of log bail amounts for defendants accused of felony drug offenses and who saw any of the five most common magistrates in our dataset. The variance of the bail assignments in these cases is too wide for meaningful inference, so we do not include it in our analysis. The stochasticity of a magistrate's  assigned cash bail amounts may be instructive in its own right as a metric for understanding judicial treatment.}
		\label{logbailhist}
	\end{figure}
	
	If we consider the defendants who were assigned cash bail by one of the magistrates in our dataset and accused of similar crimes/crime severity, after normalizing their feature vectors and selecting a subset of these defendants whose euclidean distance is less than $0.05$, we obtain a set of defendants who we would expect to have comparable bail amounts. Figure \ref{logbailhist} presents a histogram of this subset of defendants with felony drug offenses for the five most active magistrates in our dataset. Ideally, we would like to perform a regression with a bail indicator in order to approximate counterfactual effects, similarly to our approach to approximating FTA probabilities and bail posting probabilities. However, the variance of the log bail amounts is too large for meaningful estimations of counterfactuals. This tells us that bail amount assignments are far too stochastic for the granularity of data that we use here. Instead we rely on the colinearity of our other features with the log bail amounts in order to reasonably estimate the counterfactual bail posting and failure to appear probabilities without directly including the amounts as covariates.	

\section{Data Collection and Benchmark Tests}

Our dataset of cash bail decisions makes use of the publicly available Unified Judicial System of Pennsylvania's Web Portal, which records all court dockets for cases in the state of Pennsylvania. The website provides a pdf for each docket and pdf summary file for each defendant's criminal history. The Pennsylvania judiciary also includes a separate API, in which users can retrieve json formatted files of case information.  We use text-extraction tools in order to retrieve an individual's history from the summary pdf and combine this with the json docket information.

We performed this process on $33,566$ records from the Court of Common Pleas in Philadelphia, $66,982$ from the court of Common Pleas in Allegheny county, with a total set of $89,751$ cases from January 1, 2016 to December 31, 2019 over these two counties.
	
	From each docket-summary pair, we extract a feature vector for each case that consists of:
	\begin{enumerate}
		\item \textbf{Magistrate} Categorical identifier for which magistrate set the defendant's bail.
		\item \textbf{Race}: Binary variable, White/Black.
		\item \textbf{Sex}: Binary variable, Male/Female.
		\item \textbf{Age}: Defendant's age when the case was filed
		\item \textbf{Lead Offense Type}: Categorical identifier of the defendant's lead offense (eg. First/Second/Third Degree Felony or Misdemeanor).
		\item \textbf{Lead Offense Description}: Categorical identifier for the lead offense based on the lead offense's statute number: Offenses against Public Administration, Offenses against Property, Offenses Involving Danger to the Person, Drug Offenses, Inchoate Crimes, Vehicle Offenses, Driving after Imbibing Alcohol or Utilizing Drugs, Miscellaneous Offenses 
		\item \textbf{Number of charges}: Number of charges brought against the defendant in the current case
		\item \textbf{Attorney Type}: Categorical variable, Public Defense Attorney, Private Defense Attorney, Attorney Waived.
		\item \textbf{Number of Prior Felonies}: Categorical variable of prior felony convictions; zero prior convictions, one to two prior convictions, three to five prior convictions, six to nine prior convictions, ten or more convictions. 
		\item \textbf{Number of Prior Misdemeanors}: Categorical variable of prior misdemeanors; zero prior convictions, one to two prior convictions, three to five prior convictions, six to nine prior convictions, ten or more convictions. 
		\item \textbf{Bail Status}: Categorical variable of the status of the defendant's bail; Bail Set, Bail Posted, Bail Forfeited, Bail Revoked. Anyone who has not had their bail status changed from "Bail Set" is treated as being detained pretrial. Individuals whose status is, "Bail Posted", "Bail Forfeited", and "Bail Revoked" are all treated as having posted bail and been released pretrial.
		\item \textbf{Bail Type}: Categorical variable of the defendant's bail type; Monetary Bail, Unsecured Bail, Non-monetary Bail, Nominal Bail. We only consider the bail set at an initial hearing and not subsequent changes to bail types and amounts
		\item \textbf{Fail to Appear}: Binary variable describing whether or not the defendant did not appear on their court date. 
		\item \textbf{Census Based Confounders}: Each docket contains the defendant's zip code or city of residence.  We try to minimize confounders such as socioeconomic status or education level by including the following for each defendant based on the US Census 2018 American Community Survey 5-Year census estimate for their zip code of residence: 
		\begin{itemize}
			\item Median Income
			\item High School Graduation Rate 
			\item College Graduation Rate
			\item Poverty Rate
			\item Employment Rate
			\item Median Age
		\end{itemize}
	\end{enumerate}

	\subsection{Pennsylvania's Clean Slate Law}

Signed in 2018, Pennsylvania has implemented the first-of-its-kind ``Clean Slate'' law, which begun to seal millions of criminal records in an effort to allow Pennsylvania residents who have not been found guilty of serious crimes (eg. violence, sexual assault, homicide, etc.) to more easily continue with their lives. This law seals records so that they do not show up in employer background searches nor in Pennsylvania's online docket database. It allows persons whose cases have been sealed to act as if their offense had never occurred. 
		
	As a result of this law, we base our analysis on cases stemming from the Court of Common Pleas, rather than Municipal Courts, as the Court of Common Pleas sees more serious crimes and has fewer cases which are not sealed under the ``Clean Slate'' law. This allows a more consistent measurement among court decisions and will hopefully minimize the effects of covariate shift from before to after the "Clean Slate" law.

\subsection{Initial Data Statistics}
		Again, the data used in this analysis consists of court filings from Allegheny County, Pennsylvania and Philadelphia County, Pennsylvania from $2016$ to $2019$ in the Court of Common Pleas and Magisterial District Courts. The Court of Common Pleas hears major civil and criminal cases, as opposed to the Municipal Courts. 
		After removing cases with either missing or ignored features, our dataset's descriptive statistics are given below.

	\begin{center}
		 \begin{tabular}{|c|c|c|}
		 	\hline
		 	\multicolumn{3}{|c|}{Descriptive Statistics} \\
		 	\hline
		 	& Allegheny & Philadelphia \\
		 	\hline
		 	N & $35345$ & $20072$ \\
		 	Black & $0.412$ & $0.653$ \\
		 	White & $0.587$ & $0.346$ \\
		 	Male & $0.738$ & $0.869$ \\
		 	Female & $0.262$ & $0.130$ \\
		 	Public Defender & $0.509$ & $0.819$ \\
		 	Private Defender & $0.361$ & $0.181$ \\
		 	Defender Waived & $0.128$ & $0.000$ \\
		 	Felony Defendant & $0.323$ & $0.869$ \\
		 	Misdemeanor Defendants & $0.603$ & $0.068$ \\
		 	Assigned Cash Bail & $0.338$ & $0.768$ \\
		 	Posted Bail & $0.562$ & $0.477$ \\
		 	Failed to Appear (Released) & $0.047$ & $0.061$ \\
		 	Failed To Appear (Detained) & $0.003$ & $0.011$ \\
		 	\hline
		 \end{tabular}
	\end{center}
	
\subsection{Pretrial Failure}

	In the dataset used here, the number of cases in which defendants miss their court dates ($0.047-0.061$) is noticeably lower than other analyses of cash bail which have failure to appear rates of: $0.258$ (Philadelphia only; Release pretrial) \cite{arnold2018racial}, $0.179$ (Philadelphia and Miami Dade County; Released pretrial) \cite{dobbie2018effects}, $0.121$ (Philadelphia and Miami Dade County; Detained pretrial) \cite{dobbie2018effects}, $0.17$ (Philadelphia only; eligible under Philadelphia DA's No-Cash-Bail policy) \cite{ouss2020bail},  $0.06$ (Philadelphia only; ineligible under Philadelphia DA's No-Cash-Bail policy) \cite{ouss2020bail}. However, for the years $2016, 2017, 2018$, the average rate of missed court appearance for the Common Pleas Court in Philadelphia was $0.043$ \cite{melamed2019philly}. From our review, there is no clear and consistent notice that a defendant did not appear at their court date within the publicly available dockets. We flag a defendant as having failed to appear at their court date if one of two cases is satisfied within the docket: 1) if the related docket both lists their bail as having been revoked/forfeited and the reason for being revoked explicitly lists that either the defendant did not appear or that a bench warrant was put out 2) if a bench warrant was authorized and if the comments within the case registry entry explictly states that the defendant did not appear.

	Failure to appear rates are only a single aspect of pretrial failure. Defendants may fail their pretrial release conditions by being arrested for a new crime while waiting for their trials. Again, the dockets do not provide a clear and consistent notice that an individual was arrested while awaiting trial. While the court summaries do provide a view of all arrests for an individual, in many cases, the same incident was recorded multiple times, and it may also be possible that an individual was arrested in other jurisdictions, although this case is likely an uncommon occurrence over all cases. Without the ability to cross-reference our dataset with prison records, we are unable to confidently determine if an individual was arrested again pretrial. As a result, we focus on a more limited view of pretrial failure by only considering whether individuals missed their court date.
	
%

\subsection{Additional Data Challenges}
There were several other features that we would have liked to include, yet ultimately decided to remove from our dataset. In future bail analyses, it may be worth considering how some of these features play into the cost model. Such features include a) whether or not a defendant had violated their probation when they were arrested; and b) the time since a defendant's most recent arrest.

For the former, the dockets are unclear on whether probation was violated when the defendant was arrested. Some cases reference a violation of probation hearing, yet we are not confident that the occurrence of this hearing should be the sole indicator of whether a defendant was in violation of their probation. Concerning the latter, we are unable to retrieve prison data for each defendant, so there are likely cases where arrest dates would not differentiate between someone who was released 20 years ago and only recently rearrested, and someone who was imprisoned for 20 years and immediately rearrested. While both of these are highly consequential in determining bail, including this information would at best give us an extremely noisy estimate of the true features.

\subsection{Considerations on Data Release}
The data used in this analysis has been released under the following github repository: \texttt{https://github.com/jnwilliams/padockets}. In releasing the full data, we have consulted with members of the Pennsylvania ACLU on the necessary protections for the individuals represented within. Following their recommendations, we have anonymized the data by a) removing any reference to the referred docket identifiers; b) replacing magistrate names with a numerical identifier; c) replacing exact docket filing/arrest dates with only the month and year; d) removing specific ages of defendants and instead providing an age range over 5 years, beginning from 18 years of age; e) replacing the median income for the defendant's zip code with an income range of \$5,000; f) replacing the statistics for an individuals zip code (ie. high school and college graduation rate, poverty rate, unemployment rate)  with a range of 5\% (eg. poverty rate of 7\% is presented as 5-10\%).

We feel that releasing the full dataset may be beneficial for future analyses, however in making this data public, our goal was to make retrieving information for any particular case within this dataset no easier to obtain here, than by searching through the Unified Judicial System of Pennsylvania's online portal.
 

\section{Modeling Results and Analysis}
		We estimate the posterior of our random variables using PyMC4 Pre-Release, a probabilistic modeling language built on top of TensorFlow Probability \cite{tensorflow2015-whitepaper}. PyMC4 uses the No-U-Turn sampler (NUTS), a variant of Hamiltonian Monte Carlo in order to sample from the posterior. 
	
\begin{figure}
			\centering
			\includegraphics[scale=0.17]{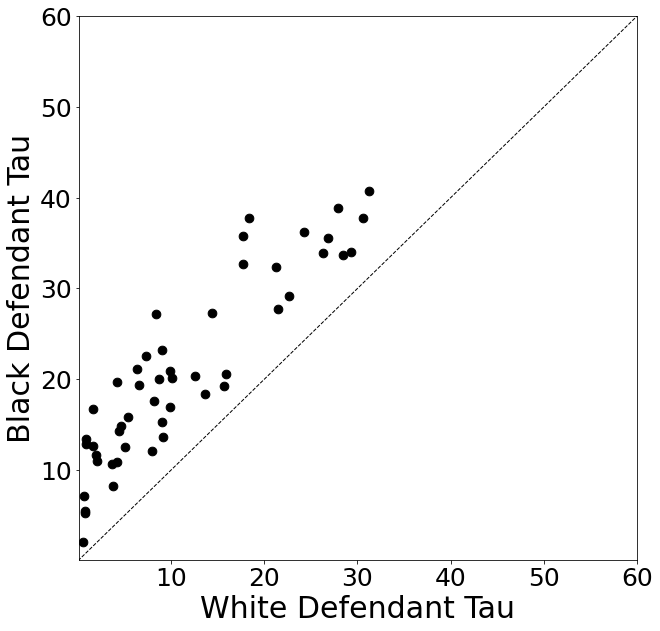}
			\includegraphics[scale=0.17]{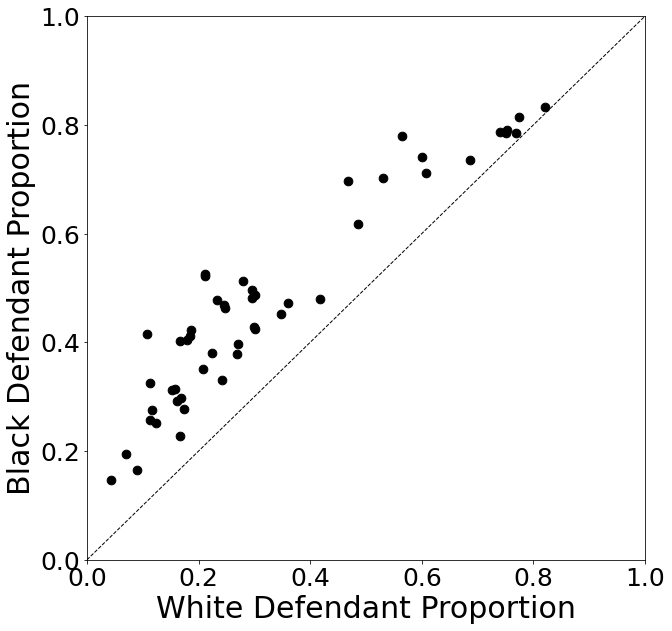}
			\includegraphics[scale=0.17]{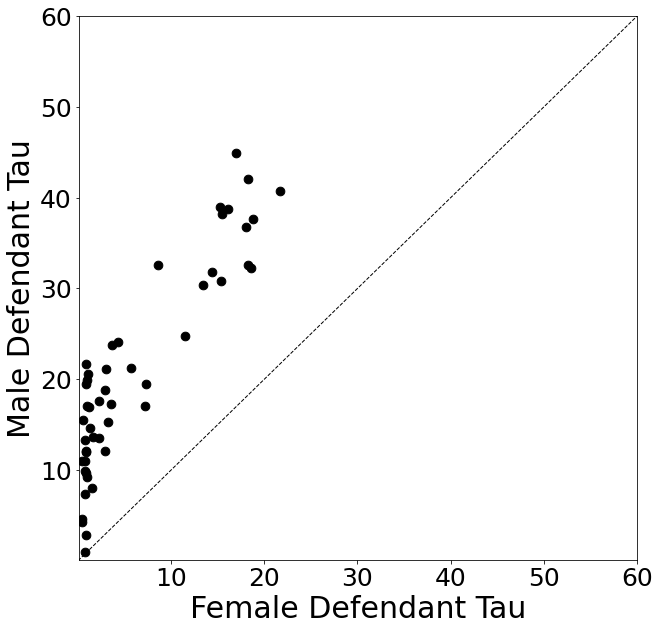}
			\includegraphics[scale=0.17]{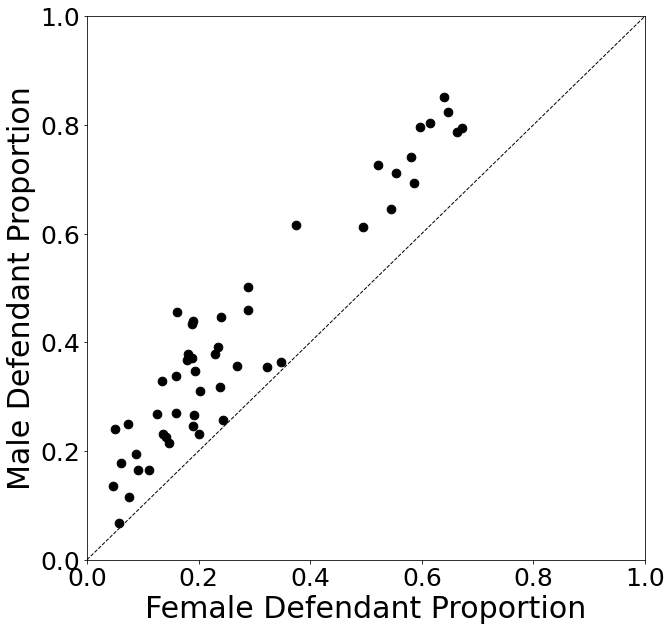}
			\includegraphics[scale=0.17]{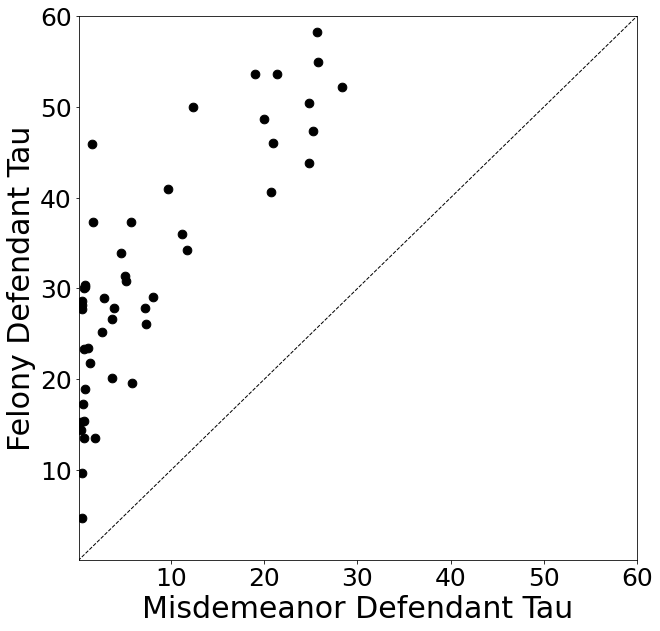}
			\includegraphics[scale=0.17]{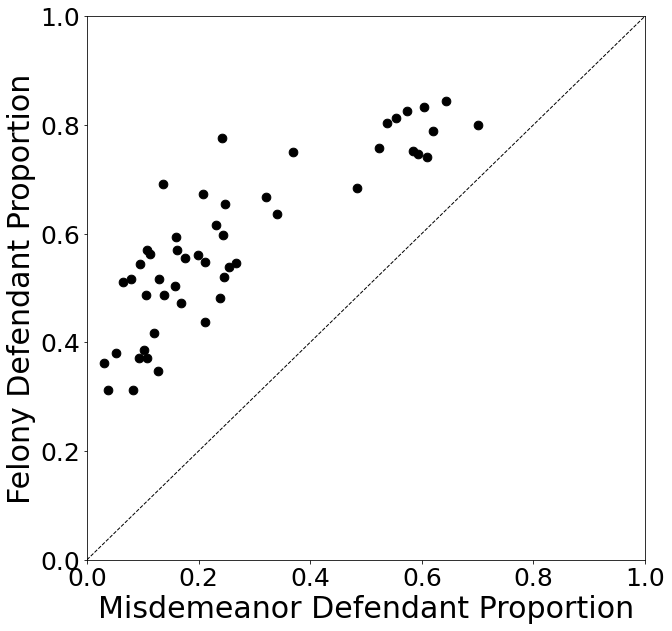}
			\caption{Comparison of computed $\tau$ values compared to the bail assignment rate for the most common $50$ magistrates. (Top Left) $\tau_{Black}$ vs $\tau_{White}$ for each magistrate. (Top Right) Proportion of Black defendants assigned cash bail vs proportion of White defendants assigned cash bail for each magistrate. (Middle Left) $\tau_{Male}$ vs $\tau_{Female}$ for each magistrate. (Middle Right) Proportion of Male defendants assigned cash bail vs proportion of Female defendants assigned cash bail for each magistrate. (Bottom Left) $\tau_{Felony}$ vs $\tau_{Misdemeanor}$ for each magistrate. (Bottom Right) Proportion of Felony defendants assigned cash bail vs proportion of Misdemeanor defendants assigned cash bail for each magistrate. }
			\label{tau}
		\end{figure}			
		
\subsection{Hamiltonian Monte Carlo Sampling}
	In high-dimensional spaces, commonly used techniques for sampling from the posterior distribution, such as the Metropolis-Hastings Algorithm or Gibbs Sampling, are susceptible to getting stuck in pathological regions of high curvature. In these spaces, the volume of the neighborhood around the mode of a distribution vanishes as the dimensionality increases, while the neighborhood away from the mode has an increasingly large volume, yet vanishing probabilities. As a result, both of these neighborhoods have a negligible effect on the expectation. 
	In such high-dimensional spaces, exploration of the typical set can lead to the discovery of small regions of high curvature.	As Monte Carlo Integration recovers a distribution's expectation asymptotically, sampling procedures tend to compensate for being unable to explore such pathological regions by spending large amounts of time exploring them when they get the opportunity. This long period of exploration causes MCMC to fail, as the sampled mean approaches the expectation within this region of high curvature rather than the expectation of the full distribution.
		
	Hamiltonian Monte Carlo Sampling avoids this issue via a more effective sampling procedure that numerically integrates along the gradient of the log-posterior in order to oscillate around the typical set, rather than perform a random walk. See \cite{betancourt2017conceptual}, for a full introduction to Hamiltonian Monte Carlo sampling.  Following standard practice, our sampling procedure consists of 2 phases, a burn-in phase, in which we sample from a random starting point until the posterior samples move toward the typical set. After this pre-set number of burn-in steps, the final posterior sampling process begins so as to give us our final parameters. In order to ensure that the posterior is not multimodal, it is common to perform this process over multiple Markov chains. 
	In this study, we found that $5$ chains, $500$ burn in steps and $1,500$ sampling steps were sufficient for all chains to converge.
		
\subsection{Results from latent variable model}
		
		\begin{figure}
			\centering
			\includegraphics[scale=0.2]{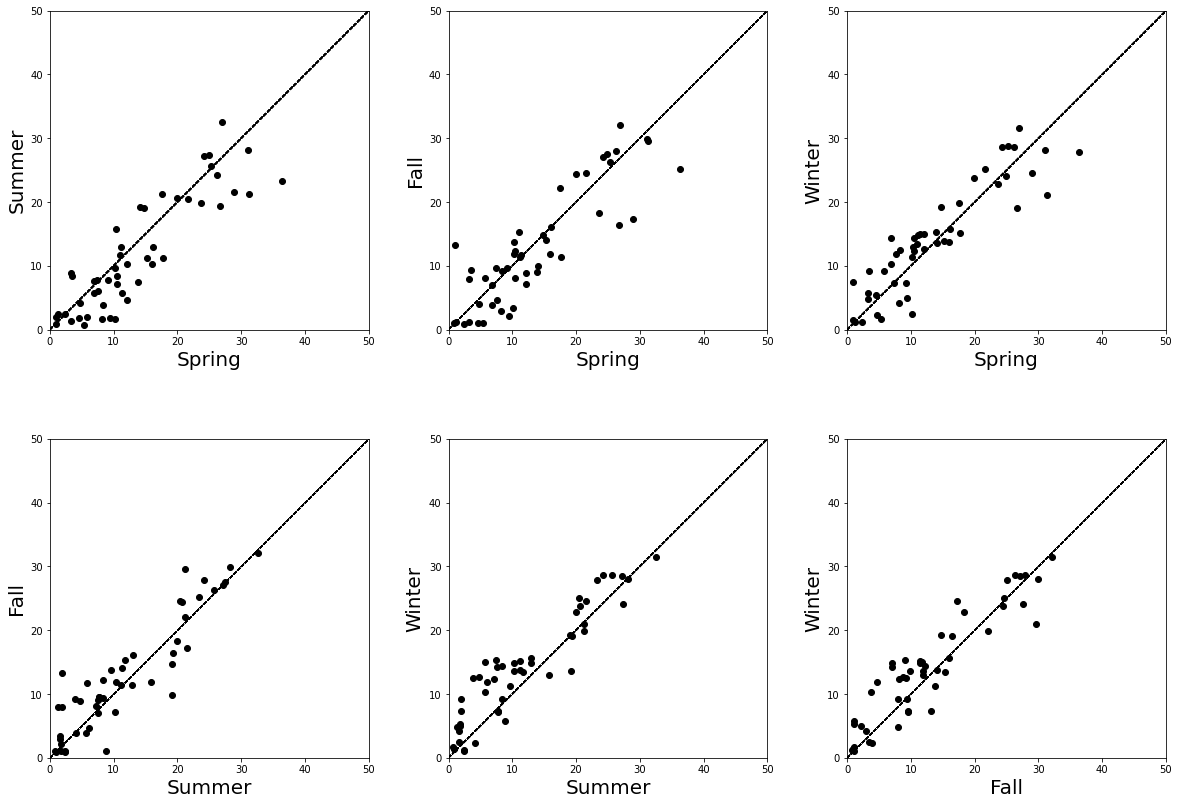}
			\caption{ Difference in $\tau$ among seasons}
			\label{season_tau}
		\end{figure}
		
		Based on the average estimated $\tau$ produced when estimating the parameters of our model, Figure \ref{tau} shows that nearly every magistrate has a striking difference in the costs of missing a court date for Black defendants as opposed to White defendants. Recall that our model defines $\tau$ as the ratio of the cost of missing a court date to the cost of being unable to post cash bail, so a greater $\tau$ implies that a judge is more willing to assign cash bail to a group in a way that minimizes the risk of missing their court date, regardless of an individual's ability to post bail. While a few magistrates have $\tau$ less then $1$ for white defendants (implying that they want to make sure that these defendants are released pretrial) no magistrate has $\tau$ less than $1$ for black defendants. Most magistrates place an outstandingly high emphasis on a defendant missing their court date for all defendants, but this trend is exacerbated based on racial differences, which is in line with the higher rate at which these magistrates assign cash bail to Black defendants.
		
		Over these magistrates, the cash bail assignment rate sits at an average $75\%$ decrease in the likelihood of being assigned cash bail when comparing their treatment of black defendants as opposed to white defendants. While such an change is indicative of a direction of bias, the results based on $\tau$ give a more actionable response to this behavior. The majority of judges have high $\tau$ values for Black defendants, with the median value being $19.51$, whereas White defendants have a median $\tau$ $8.82$. In this case, it is much more clear that the action for improving pretrial release rates requires addressing magistrate bail assignments or assigned bail amounts for Black defendants.
		
		Similarly when comparing $\tau$ for men and women, we see a similar trend in which most judges assign cash bail with less regard for male defendant's ability to pay. In this case, there may be confounding reasons why a magistrate may be more willing to ensure that women are able to be released pretrial, such as child care.  We see a similar trend in the greater proportion in which male defendants are assigned cash bail, however, as in the case or racial differences, we are able to better pinpoint the specific judges who place lower cost on some individuals being unable to pay their bail than others. This again can be a more targeted mechanism for addressing the the cash bail assignment habits of the magistrates. 
		
		We also consider two additional settings in which we want to determine the disparate treatment of those defendants accused of felonies compared misdemeanors and the treatment of defendants with respect to the season (ie. Spring, Summer, Fall, Winter). Both of these act as validation for the simple model here. As conventional wisdom would suggest that those defendants accused of felonies are significantly more likely to be assigned cash bail than defendants accused of misdemeanors, and that the treatment of defendants should remain relatively constant regardless of the season, we expect that this analysis would show similar effects.

		For those defendants accused of felonies, magistrates are much more concerned with ensuring that these defendants appear at their trial date, hence they are significantly more likely to assign cash bail. Many magistrates assign cash bail so as to ensure that those who are accused of misdemeanors are able to be released pretrial. However, still we show that there are some who are relatively unconcerned with defendants being able to post bail, regardless of offense severity. 
		
		Figure \ref{season_tau} provides the comparison of $\tau$ based on seasons, where Spring groups together all cases between March to May of any year, Summer groups together all cases between June to August, Fall takes place from September to November, and Winter is December to January. In this additional validation step, comparisons over $\tau$ should not be susceptible to groupings in which we expect cash bail assignments to be invariant, such as seasonal differences. Here, we see this effect; there is little difference in $\tau$ among each season. Our model is able to capture conventional wisdom on how cash bail assignments are done regarding offense severity and groupings in which these cash bail assignments should be invariant.
		

	\subsection{Posterior predictive checks}
	
		 \begin{figure}
	 	\includegraphics[scale=0.225]{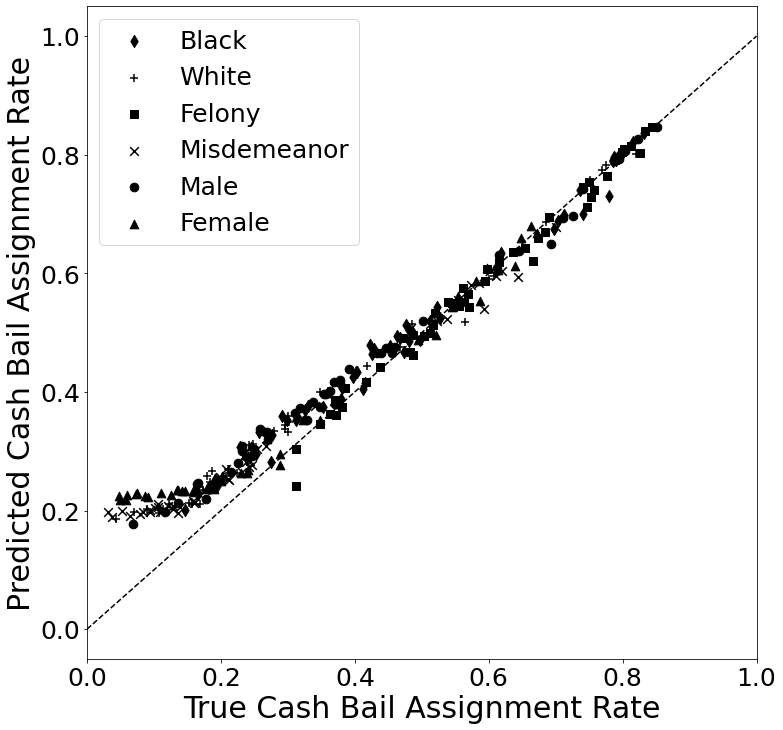} 			
	 	\caption{Average difference of the bail assignment rate via the posterior draws compared to the true bail assignment rate with respect to each subgroup. }
	 	\label{ppc}
	 \end{figure}
	 
	As a final robustness step of our model, we perform posterior predictive checks on our observed variables. Such checks use the sampled joint posterior distribution to generate new outcome variables for every case, telling us, under our posterior, whether or not a specific defendant will be assigned cash bail. We present these results in Figure \ref{ppc}. Our model shows that among the groups (gender, race, offense severity),we are able to match the expected cash bail assignment rates of each magistrate, giving credence that the approximated $\tau$ is able to act as a descriptor of magistrate behavior. 
	
	We do see a significant drop-off in accuracy for magistrates who assign cash bail in less than $20\%$ of cases. This is likely due to a limitation in our model. Based on the cost model shown in equation \eqref{taumodel}, the co-domain of the difference in magisterial cost of assigning cash bail is limited to the range $[ -1, \inf )$. The re-scaling done in equation \eqref{sigmarescale} is able to successfully make this a probability distribution, however, we still have significantly less flexibility in cases in which magistrates are unlikely to assign cash bail. This likely results in our model being unable to create a strong descriptor among these magistrates.	
	
	While our model's efficacy drops off for magistrates who have low rates of cash bail assignment, in most circumstances, it is unlikely that magistrates who assign cash bail in less than $20\%$ of cases are abusing the cash bail system to the extent that magistrates with higher cash bail assignment rate may. As such, this model may be useful to understand the behavior of problematic magistrates, whereas other methods may be beneficial for less problematic magistrates.

\section{Discussion and Conclusions}

In this paper we have revisited the infra-marginality problem for cash bail assignments in pretrial decisions. There has been a significant amount of work in understanding why judges assign cash bail at disparate rates among groups, and when such analyses compare the rate of observable outcomes such as bail assignment, recidivism, or pretrial failure, these outcomes may in turn be influenced by the unobserved features that cause an incomplete view of the system. Much work attempts to mitigate these effects by including measures of judge leniency in analyses on the influence cash bail decisions, however, these measures may also be influenced by exogenous variables.

We address this problem via Bayesian modeling, so as to focus on understanding how well judge decision making matches both the believed purposes and the legal requirements for setting cash bail. This method may be useful in providing a more nuanced view of why judges assign cash bail to certain groups at a higher rate than others. In this work, we show not only that in the Court of Common Pleas, magistrates universally assign a lower cost to Black defendants being unable to be released pretrial as opposed to White defendants, but that these estimated costs are able to avoid the infra-marginality problem. Hence, these costs are more representative of real world effects. By presenting this view in which we can directly infer how judges value an individuals ability to pay cash bail, we can better address disparities in the cash bail system, and more closely target the exact mechanisms by which these disparate treatments occur.

While the results presented here suggest specific values for the underlying beliefs for each magistrate, due to the limitations here, it would be beneficial to repeat this analysis with a dataset that includes more information on pretrial failure, such as arrests for other crimes while released pretrial. This is especially important, because pretrial release decisions are based on more than missing a court date alone. For example, by replacing $\pr{ \mathrm{fta} | \mathrm{cash\ bail} }$ with $\pr{ \mathrm{pretrial\ failure} | \mathrm{cash\ bail} }$ alone may provide a fuller view of magistrate decision-making than we present here. Despite this, we believe that the results presented here are an important study that should be built on further, and that an additional method for addressing the problem of infra-marginality in cash bail decisions is a useful contribution for the broader algorithmic community.


\section*{Acknowledgements}

We thank Jessica Li and Nyssa Taylor from the American Civil Liberties Union of Pennsylvania for useful discussions regarding our data collection efforts and in helping us gain a better understanding of the cash bail system. 

\bibliographystyle{ACM-Reference-Format}
\bibliography{bibliography}

\appendix

\end{document}